\documentclass[12pt]{article}
\usepackage{amsfonts}
\usepackage[mathscr]{eucal}

\newcommand{\be}{\begin{equation}}\newcommand{\ee}{\end{equation}}
\newcommand{\bea}{\begin{eqnarray}}
\newcommand{\eea}{\end{eqnarray}}

\newcommand{\lb}{\label}
\newcommand{\p}[1]{(\ref{#1})}

\begin{document}

\begin{titlepage}
\begin{flushright}
ITP-UH-08/09
\end{flushright}

\vspace*{1cm}

\renewcommand{\thefootnote}{\dag}
\begin{center}

{\LARGE\bf OSp(4$|$2) Superconformal Mechanics }

\vspace{2cm}
\renewcommand{\thefootnote}{\star}

{\large\bf S.~Fedoruk}${}^1$,\,\,\, {\large\bf E.~Ivanov}${}^1$,\,\,\,
{\large\bf
O.~Lechtenfeld}${}^2$ \vspace{1cm}

${}^1${\it Bogoliubov  Laboratory of Theoretical Physics, JINR,}\\
{\it 141980 Dubna, Moscow region, Russia} \\
\vspace{0.1cm}

{\tt fedoruk,eivanov@theor.jinr.ru}\\
\vspace{0.5cm}

${}^2${\it Institut f\"ur Theoretische Physik,
Leibniz Universit\"at Hannover,}\\
{\it Appelstra{\ss}e 2, D-30167 Hannover, Germany} \\
\vspace{0.1cm}

{\tt lechtenf@itp.uni-hannover.de}\\
\vspace{0.3cm} \setcounter{footnote}{0}

\end{center}
\vspace{0.2cm} \vskip 0.6truecm  \nopagebreak

\begin{abstract}
A new superconformal mechanics with OSp(4$|$2) symmetry is obtained
by gauging the U(1) isometry of a superfield model. It is the
one-particle case of the new ${\cal N}{=}4$ super Calogero model
recently proposed in {\tt arXiv:0812.4276~[hep-th]}. Classical and
quantum generators of the $osp(4|2)$ superalgebra are constructed on
physical states. As opposed to other realizations of ${\cal N}{=}4$
superconformal algebras, all supertranslation generators are linear
in the odd variables, similarly to the ${\cal N}{=}2$ case. The
bosonic sector of the component action is standard one-particle
(dilatonic) conformal mechanics accompanied by an ${\rm SU}(2)/{\rm
U}(1)$ Wess-Zumino term, which gives rise to a fuzzy sphere upon
quantization. The strength of the conformal potential is quantized.
\end{abstract}

\vspace{1cm}
\bigskip
\noindent PACS: 03.65.-w, 04.60.Ds, 04.70.Bw, 11.30.Pb

\smallskip
\noindent Keywords: Superconformal symmetry, black holes,
superfields

\newpage

\end{titlepage}

\setcounter{footnote}{0}

\setcounter{equation}0
\section{Introduction}

The stable interest in conformal mechanics
\cite{AFF,Ja,IKLe,CDKKTP,GT,MS,Pa,Plyu,IKN}
and its various superconformal extensions
\cite{AP,FR,IKL,CDKKTP,GT,AIPT,Wyl,MS,Pa,IKN,BGIK,IKLech,IL,IKLecht,
GalLe,Gal,BKSS,KriLePol}
is caused by two closely connected reasons. First, these models
describe (super)particles moving in near-horizon (AdS) geometries
of black-hole solutions of supergravities in diverse dimensions
and so bear an intimate relation to the AdS/CFT correspondence.
Second, they are one-particle prototypes of many-particle
$d{=}1$ integrable (super)conformal systems of the Calogero type,
which are the object of numerous studies.
The search for new models of this kind and their implications in the
areas just mentioned present interesting venues for study.

It has been proposed in~\cite{CDKKTP} that the radial motion of a
massive charged particle near the horizon of an extremal
Reissner--Nordstr\"om (RN) black hole is described by conformal
mechanics~\cite{AFF}. The target variable of this conformal
mechanics is the ${\rm AdS}_2$ radial coordinate as part of the
${\rm AdS}_2\times S^2$ background. The latter is the bosonic body of
the maximally supersymmetric near--horizon extremal RN~(Reissner-Nordstr\"om)
solution of ${\cal N}{=}2$ $D{=}4$ supergravity~\cite{KR,CDKKTP}, with the
full isometry supergroup ${\rm SU}(1,1|2)$. Based on this observation,
it was suggested in~\cite{CDKKTP} that the ${\rm SU}(1,1|2)$ ${\cal N}{=}4$
superconformal mechanics describes the full dynamics of a superparticle
in the near--horizon geometry of extremal RN~black holes.

${\rm SU}(1,1|2)$ superconformal mechanics
was constructed and investigated more than twenty years ago in~\cite{IKL}
in the framework of the nonlinear realizations approach.
In~\cite{AIPT}, some of the results of~\cite{IKL} were rediscovered and
transported into the modern black-hole and AdS/CFT context.
In~\cite{GT}, it was then argued that an $n$--particle generalization of
the ${\rm SU}(1,1|2)$ superconformal mechanics,
in the form of a superconformal Calogero model, in the large--$n$ limit
provides a microscopic description of {\it multiple\/} extremal RN black
holes in the near--horizon limit.
Further evidence in favor of the proposal of~\cite{CDKKTP} was adduced
in~\cite{IKN,BGIK}, where a canonical transformation was found to link
the radial motion of a (super)particle on ${\rm AdS}_2\times S^2$
as bosonic background with ${\cal N}{=}0$, ${\cal N}{=}2$~\cite{IKN} and
${\cal N}{=}4$~\cite{BGIK} superconformal mechanics.

There are good reasons to look beyond ${\rm SU}(1,1|2)$ to the most general
${\cal N}{=}4$ superconformal group in one dimension, which is the exceptional
one--parameter supergroup $D(2,1;\alpha)$~\cite{Sorba}. It reduces to
${\rm SU(1,1|2)}{\subset\!\!\!\!\!\!\times}{\rm SU(2)}$ at $\alpha{=}0$
and $\alpha{=}-1$.
In fact, the isometry supergroup of a near--horizon $M$--brane solution
of $D{=}11$ supergravity was determined as $D(2,1;\alpha){\times}D(2,1;\alpha)$
\cite{GMT-2}, and $D(2,1;\alpha)$ is physically realized for any value of
the parameter~$\alpha$ in the near--horizon $M$--theory solutions~\cite{Town}.

The general one--dimensional sigma model with $D(2,1;\alpha)$ supersymmetry,
in terms of ${\cal N}{=}1$ superfields, was applied in~\cite{Pa} to the
non--relativistic spinning particle propagating in a curved background
augmented with a magnetic field and a scalar potential.
$D(2,1;\alpha)$ superconformal mechanics was also constructed in~\cite{IKLech}
in the nonlinear realizations superfield framework.
Described by the ${\bf (3,4,1)}$ off-shell ${\cal N}{=}4$ supermultiplet,
this model contains in its bosonic sector three fields, which stand for the
dilaton and for the coordinates of the coset $S^2\simeq{\rm SU}(2)/{\rm U}(1)$,
thus governing a particle moving on~${\rm AdS}_2\times S^2$.
Furthermore, with the help of a special canonical transformation, a recent
paper~\cite{Gal} established a connection between the model of~\cite{IKLech}
with $D(2,1;-1)\simeq{\rm SU}(1,1|2){\subset\!\!\!\!\!\!\times}{\rm SU}(2)$
invariance and a particle propagating near the horizon of extremal RN black
hole with magnetic charge.

In the present paper we construct and examine a new type of ${\cal N}{=}4$
superconformal mechanics model, which is invariant under the
supergroup $D(2,1;-\frac12)\simeq{\rm OSp}(4|2)\,$.
We note that ${\rm OSp}(4|2)$ is distinguished among all ${\cal N}{=}4$
supergroups $D(2,1;\alpha)$ because its coset superspace
${\rm OSp}(4|2)/[{\rm SO}(1,1)\times{\rm SO}(2)\times{\rm SU}(2)]$
is the only superextension of ${\rm AdS}_2\times S^2$ which admits
a superconformally flat supervielbein and superconnections, as opposed
to the more conventional coset superspace
${\rm SU}(1,1|2)/[{\rm SO}(1,1)\times{\rm SO}(2)]$~\cite{BILS}.

Our new ${\rm OSp}(4|2)$ mechanics arises as the
$n{=}1$ case of an ${\cal N}{=}4$ supersymmetric generalization of the
$A_{n-1}$ Calogero system proposed recently in~\cite{FIL}, and it
radically differs from the model of~\cite{IKLech}. For one, our model is
defined by a {\it reducible\/} $D(2,1;-\frac12)$ representation, namely it is
a coupled system of one ${\bf (1,4,3)}$ multiplet and one ${\bf (4,4,0)}$
multiplet, both presented by appropriate bosonic superfields. In the action,
the ${\bf (4,4,0)}$ multiplet is described by a pure superfield Wess-Zumino
term, without standard kinetic term. Furthermore, our model posesses a gauged
${\rm U}(1)$ symmetry, ensured by a non-propagating gauge multiplet. After
fixing this ${\rm U}(1)$ in a manifestly ${\cal N}{=}4$ supersymmetric way,
the ${\bf (4,4,0)}$ Wess-Zumino multiplet turns into a ${\bf (3,4,1)}$
multiplet, which superconformally couples to the ${\bf (1,4,3)}$ superfield.
Alternatively, a Wess-Zumino gauge choice may be more suitable for
analyzing the component structure and for its quantization.

In the next two sections we give a general description of the model, first in
superfields and then in component fields. Quantization is performed in Sect.~4.
We employ the harmonic framework of~\cite{GIOS,IL}.
Thus, from the very beginning, the super worldline is extended by
${\rm SU}(2)/{\rm U}(1)$ harmonics.
After eliminating auxiliary fields in the component action, we obtain
harmonic--like fields also in the target space. The action for these fields
is only of first order in time derivatives, hence get quantized to pure spin
(or ``isospin'') degrees of freedom. Thus, starting from a theory with
worldline harmonic variables, we arrive at a sort of harmonic target
superspace. The corresponding wave functions are irreducible ${\rm SU}(2)$
multispinors, in contradistinction to ordinary conformal or superconformal
quantum mechanics \cite{AFF,AP,FR,IKL,AIPT} where spin is solely due to the
fermionic fields and disappears in the bosonic limit. Here instead, the bosonic
quantum sector may be interpreted as a direct product of standard quantum
conformal mechanics~\cite{AFF} with a {\it fuzzy sphere\/}~\cite{Mad},
which appears by virtue of the $S^2$ Wess-Zumino term.

\setcounter{equation}0
\section{Superfield setup}

A natural arena for ${\cal N}{=}4, d{=}1$ supersymmetric theories is the
${\cal
N}{=}4, d{=}1$
superspace~\cite{IKL}
\be
(t,\theta_i, \bar\theta^i)\,, \qquad \bar\theta^i=(\overline{\theta_i})\,,
\quad (i
= 1, 2)\,. \lb{Rss}
\ee
The corresponding spinor covariant derivatives have the form
$$
D^i=\frac{\partial}{\partial\theta_i}+i\bar\theta^i \partial_t\,,\qquad \bar
D_i=\frac{\partial}{\partial\bar\theta^i}+i \theta_i \partial_t = -
\overline{(D^i)}\,.
$$
The full R-symmetry (automorphism) group of \p{Rss} is ${\rm SO}(4)_R\,$.
One of the
two ${\rm SU}(2)$ factors
of the latter acts on the doublet indices $i$ and will be denoted ${\rm
SU}(2)_R\,$.
The second ${\rm SU}(2)$ mixes
$\theta_i$ with their complex conjugates and is not manifest in the
considered
approach.

Off-shell ${\cal N}{=}4, d{=}1$ supermultiplets admit a concise
formulation in
the harmonic superspace (HSS)~\cite{IKLech}, an extension of \p{Rss} by
the harmonic
coordinates $u^\pm_i$:
\be
(t,\theta^\pm, \bar\theta^\pm, u_i^\pm)\,, \qquad \theta^\pm=\theta^i
u_i^\pm \,,
\qquad
\bar\theta^\pm=\bar\theta^i u_i^\pm\,,\qquad u^{+i}u_i^-=1\,. \lb{HSS}
\ee
The commuting ${\rm{SU}}(2)$ spinors $u_i^\pm$ parametrize the 2-sphere
$S^2 \sim {\rm SU}(2)_R/{\rm{U}}(1)_R$. The salient property of HSS is the
presence of
an important subspace in it, the harmonic analytic superspace (ASS) with
half of
Grassmann co-ordinates
as compared to \p{Rss} or \p{HSS}:
\be
(\zeta,u)=(t_A,\theta^+, \bar\theta^+, u_i^\pm)\,, \qquad t_A=t-i(\theta^+
\bar\theta^-
+\theta^-\bar\theta^+)\,. \lb{Ass}
\ee
It is closed under the ${\cal N}{=}4$ supersymmetry transformations. Most
of the
off-shell ${\cal N}{=}4, d{=}1$
multiplets are represented by the analytic superfields, i.e. those
``living'' on
\p{Ass}.

Spinor covariant derivatives in the analytic basis of HSS, viz. $(\zeta, u,
\theta^-, \bar\theta{}^-)$, take the form
\begin{equation}\label{D-ferm}
D^+ =\frac{\partial}{\partial\theta^-}\,,\quad \bar D^+
=-\frac{\partial}{\partial\bar\theta^-}\,,\qquad D^-
=-\frac{\partial}{\partial\theta^+}+2i\bar\theta^-\partial_{t_A}\,,\quad
\bar D^-
=\frac{\partial}{\partial\bar\theta^+}+ 2i\theta^-\partial_{t_A}\,.
\end{equation}
In the central basis \p{HSS}, the same derivatives are defined as the
projections
$D^\pm=D^i u_i^\pm$ and $\bar D^\pm=\bar D^i u_i^\pm$.
Harmonic covariant derivatives in the analytic basis read
\begin{equation}\label{D-harm}
D^{\pm\pm} =\partial^{\pm\pm} -2i\theta^\pm\bar\theta^\pm\partial_{t_A} +
\theta^\pm\frac{\partial}{\partial\theta^\mp}
+\bar\theta^\pm\frac{\partial}{\partial\bar\theta^\mp}\,.
\end{equation}
The integration measures are defined by
$$
\mu_H =dudtd^4\theta =\mu^{(-2)}_A(D^+\bar D^+)\,, \qquad
\mu^{(-2)}_A=dud\zeta^{(-2)}
=dudt_A d\theta^+d\bar\theta^+ =dudt_A(D^-\bar D^-)\,.
$$

\subsection{Action}

In~\cite{FIL}, we constructed a new ${\cal N}{=}4$ supersymmetric
extension of the $A_{n-1}$ Calogero system. Distinguishing features
of its Lagrangian are, first, the appearance of the $U(2)$ spin
generalization of the $A_{n-1}$ Calogero in its bosonic sector,
second, ${\cal N}{=}4$ superconformal invariance associated with the
supergroup $D(2,1;-\frac12)\simeq {\rm{OSp}}(4|2)$ (as opposed to
the ${\rm SU}(1,1|2)$ superconformal symmetry of the standard ${\cal
N}=4$ superextensions \cite{IKL,AIPT}) and, third, a nontrivial
coupling to the center-of-mass coordinate. All these features are
retained even in the extremal $n{=}1$ case where only the
center-of-mass coordinate is present. It develops a conformal
potential, so the $n{=}1$ case of the ${\cal N}{=}4$ Calogero model
of~\cite{FIL} amounts to a non-trivial model of ${\cal N}{=}4$
superconformal mechanics (as distinct from the new ${\cal N}{=}1,2$
models also obtained in \cite{FIL} by the same method; in them,  the
$n{=}1$ case yields a free system). Below we describe the superfield
action of this model.

It involves superfields corresponding to three off-shell ${\cal N}{=}4$
supermultiplets: ({\bf i}) the ``radial'' multiplet
({\bf 1,4,3}); ({\bf ii}) the Wess-Zumino (``isospin'') multiplet ({\bf
4,4,0});
and ({\bf iii}) the gauge (``topological'') multiplet ({\bf 0,0,0}). The
action has
the form
\begin{equation}\label{4N-gau}
S =S_{\mathscr{X}} + S_{FI} + S_{WZ}\,.
\end{equation}

First term in~(\ref{4N-gau}) is the standard free action of ({\bf 1,4,3})
multiplet
\begin{equation}\label{4N-X}
S_{\mathscr{X}} =-{\textstyle\frac{1}{2}}\int \mu_H \, \mathscr{X}^{\,2} \,,
\end{equation}
where the even real superfield $\mathscr{X}$ is subjected to the constraints
\begin{equation}  \label{cons-X-g-V}
D^{++} \,\mathscr{X}=0\,,
\end{equation}
\begin{equation}  \label{cons-X-g}
D^{+}{D}^{-} \,\mathscr{X}=0\,,\qquad
    \bar D^{+}\bar D^{-}\, \mathscr{X}=0\,,\qquad
    (D^{+}\bar D^{-} +\bar D^{+}D^{-})\, \mathscr{X}=0\,.
\end{equation}
The set of conditions \p{cons-X-g-V} and~(\ref{cons-X-g}) is equivalent to
the
standard constraints $D^iD_i \,\mathscr{X}=0$,
$\bar D_i\bar D^i \,\mathscr{X}=0$, $[D^i,\bar D_i]\, \mathscr{X}=0$ in
the central
basis \p{HSS}.

Second term in~(\ref{4N-gau}) is Fayet--Iliopoulos (FI) term
\begin{equation}\label{4N-FI}
S_{FI} ={\textstyle\frac{i}{2}}\,\,c\int \mu^{(-2)}_A \,V^{++}
\end{equation}
for the gauge supermultiplet. The even analytic
gauge superfield $V^{++}(\zeta,u)$, ${D}^{+} \,V^{++}=0$, $ \bar{D}^{+}\,
V^{++}=0\,,$ is subjected to the gauge transformations
\be
V^{++}{}' = V^{++} - D^{++}\lambda, \quad \lambda = \lambda(\zeta,
u)\,,\label{tran4-V}
\ee
which are capable to gauge away, {\it locally}, all the components from
$V^{++}$.
However, the latter contains a component which
    cannot be gauged away {\it globally}. This is the reason why this $d=1$
supermultiplet was called ``topological'' in \cite{DI}.

Last term in~(\ref{4N-gau}) is Wess--Zumino (WZ) term
\begin{equation}\label{4N-WZ}
S_{WZ} = {\textstyle\frac{1}{2}}\,\int \mu^{(-2)}_A \, \mathcal{V}\,
\bar{\mathcal{Z}}{}^+\, {\mathcal{Z}}^+\, .
\end{equation}
Here, the complex analytic superfield ${\mathcal{Z}}^+,
\bar{\mathcal{Z}}^+$ $(D^+
{\mathcal{Z}}^+ = \bar D^+ {\mathcal{Z}}^+ =
0)\,,$ is subjected to the harmonic constraints
\begin{equation}  \label{cons-Ph-g}
\mathscr{D}^{++} \,{\mathcal{Z}}^+\equiv (D^{++} + i\,V^{++})
\,{\mathcal{Z}}^+=0\,,\qquad \mathscr{D}^{++}
\,\bar{\mathcal{Z}}{}^+\equiv (D^{++} - i\,V^{++}) \,\bar{\mathcal{Z}}{}^+=0
\end{equation}
and describes a gauge-covariantized version of the ${\cal N}{=}4$
multiplet ({\bf
4,4,0}). The relevant gauge transformations
are
\be
{\mathcal{Z}}^+{}' = e^{i\lambda} {\mathcal{Z}}^+,\qquad
\bar{\mathcal{Z}}{}^+{}' =
e^{-i\lambda}\bar{\mathcal{Z}}{}^+\,.\label{tran4-Phi}
\ee

The superfield $\mathcal{V}(\zeta,u)$ in \p{4N-WZ} is a real analytic gauge
superfield (${D}^{+}
\,\mathcal{V}=\bar{D}^{+}\, \mathcal{V}=0$), which is a prepotential
solving the
constraints~(\ref{cons-X-g-V}) for $\mathscr{X}$. It is related to the
superfield
$\mathscr{X}$ in the central basis
by the harmonic integral transform~\cite{DI1}
\begin{equation}  \label{X0-V0}
\mathscr{X}(t,\theta_i,\bar\theta^i)=\int du \,\mathcal{V} \left(t_A,
\theta^+,
\bar\theta^+, u^\pm \right) \Big|_{\theta^\pm=\theta^i u^\pm_i,\,\,\,
\bar\theta^\pm=\bar\theta^i u^\pm_i}\,.
\end{equation}
The unconstrained analytic prepotential $\mathcal{V}$ has its own pregauge
freedom
\be
\delta \mathcal{V} = D^{++}\lambda^{--}\,, \quad \lambda^{--} =
\lambda^{--}(\zeta,
u)\,,\lb{ga-V0}
\ee
which can be exploited to show that $\mathcal{V}$ describes just the
multiplet
$({\bf 1, 4, 3})$
(after choosing the appropriate Wess-Zumino gauge) \cite{DI1}.
The coupling to the multiplet $({\bf 1, 4, 3})$ in \p{4N-WZ} is introduced
for
ensuring superconformal invariance.
As we shall see, upon passing to components, it gives rise to non-trivial
interactions for the physical fields.
The invariance of \p{4N-WZ} under \p{ga-V0} is ensured by the constraints
\p{cons-Ph-g}.

\subsection{Superconformal invariance}

Besides the gauge ${\rm U}(1)$ symmetry \p{tran4-V}, \p{tran4-Phi} and
pregauge
symmetry \p{ga-V0},
the action~(\ref{4N-gau}) is invariant under the rigid
${\cal N}{=}4$ superconformal symmetry $D(2,1;\alpha)$ with $\alpha=-1/2$.
All
superconformal transformations are
contained in the closure of the supertranslations and superconformal boosts.

Invariance of the action~(\ref{4N-gau}) under the supertranslations
($\bar\varepsilon^i=\overline{(\varepsilon_i)}$)
$$
\delta t =i(\theta_k\bar\varepsilon^k -\varepsilon_k\bar\theta^k),\qquad
\delta
\theta_k=\varepsilon_k, \qquad \delta \bar\theta^k=\bar\varepsilon^k
$$
is automatic because we use the ${\cal N}{=}4$ superfield approach.

The coordinate realization of the superconformal boosts of
$D(2,1;\alpha)$ \cite{IL,DI} specialized to the case of $\alpha=-1/2$ is
as follows
($\bar\eta^i=\overline{(\eta_i)}$):
\begin{equation}  \label{sc-coor-c}
\delta^\prime t=-\Lambda_0 t\,,\qquad \delta^\prime \theta_i= \eta_i t
-\Lambda_0
\theta_i\,,\qquad \delta^\prime \bar\theta^i= \bar\eta^i t -\Lambda_0
\bar\theta^i\,,
\end{equation}
\begin{equation}  \label{sc-coor-a}
\delta^\prime t_A=-2\Lambda t_A\,,\quad \delta^\prime \theta^+= \eta^+ t_A
+i\eta^-
\theta^+\bar\theta^+\,,\quad \delta^\prime \bar\theta^+= \bar\eta^+ t_A
+i\bar\eta^-
\theta^+\bar\theta^+ \,,\quad \delta^\prime u^+_i= \Lambda^{++}u^-_i\,,
\end{equation}
\begin{equation}  \label{sc-me1}
\delta^\prime (dtd^4\theta)=2\,(dtd^4\theta)\,\Lambda_0\,,\quad
\delta^\prime \mu_H=
\mu_H\left(2\Lambda +\Lambda_0\right)\,,\quad \delta^\prime \mu^{(-2)}_A=
0\,, \quad
\delta^\prime D^{++} = -\Lambda^{++}\,D^0\,,
\end{equation}
where
\begin{equation}  \label{def-La1}
\Lambda = \tilde\Lambda = i(\eta^-\bar\theta^+ - \bar\eta^-\theta^+
)\,,\qquad
\Lambda^{++}
= D^{++}\Lambda=i(\eta^+\bar\theta^+ - \bar\eta^+\theta^+ )\,, \qquad
D^{++}\Lambda^{++}=0\,,
\end{equation}
\begin{equation}  \label{def-La2}
\Lambda_0 = 2\Lambda- D^{--} \Lambda^{++} = i(\eta_k\bar\theta^k+\bar\eta^k
\theta_k)\,,
\qquad D^{++}\Lambda_0=0\,.
\end{equation}
Taking the field transformations in the form (here we use the ``passive''
interpretation of them)
\begin{equation}  \label{sc-1n}
\delta^\prime \mathscr{X}= -\Lambda_0\,\mathscr{X}\,,\qquad \delta^\prime
\mathcal{V} =
-2\Lambda\,\mathcal{V}\,,\qquad \delta^\prime {\mathcal{Z}}^+ =
\Lambda\,{\mathcal{Z}}^+\,,\qquad\delta^\prime V^{++} = 0\,,
\end{equation}
it is easy to check the invariance of the action~(\ref{4N-gau}). Note that
the
constraints~(\ref{cons-X-g-V}), (\ref{cons-X-g})
and~(\ref{cons-Ph-g}) as well as the actions~(\ref{4N-FI})
and~(\ref{4N-WZ}), are
invariant with respect to the $D(2,1;\alpha)$ transformations with an
arbitrary
$\alpha$. It is
important, that the action~(\ref{4N-WZ}) is superconformally invariant
just due to
the presence of the analytic
prepotential $\mathcal{V}\,$. The free action \p{4N-X} is invariant only
under the
supergroup $D(2,1;\alpha =-1/2) \sim {\rm OSp}(4|2)\,$
which is thus the superconformal symmetry of the full action \p{4N-gau}.

\subsection{Supersymmetric gauge}

In the next sections we will analyse the component structure of the model by
choosing the Wess-Zumino
(WZ) gauge for the superfield $V^{++}$. However, in order to clarify the
off-shell
superfield content of our model,
it is instructive to fix the underlying U(1) gauge freedom by choosing a
gauge
which preserve manifest ${\cal N}{=}4$ supersymmetry. A gauge suitable for
our
purpose was used in \cite{DI}.

To make contact with the consideration in \cite{DI}, let us combine the
superfields
${\mathcal{Z}}^+$ and $\bar {\mathcal{Z}}^+$ into a doublet of
some extra (``Pauli-G\"ursey'') SU(2)$_{PG}$ group as
\be
q^{+ a} := (\bar {\mathcal{Z}}^+, {\mathcal{Z}}^+)\,, \; a =1,2\,
\ee
and rewrite the transformation law \p{tran4-Phi} and the constraints
\p{cons-Ph-g} as
\be
\delta q^{+ a} = \lambda c^a_{\;b} q^{+ b}\,, \quad D^{++}q^{+ a} + V^{++}
c^a_{\;b}q^{+ b} = 0\,. \label{qc}
\ee
Here, the traceless constant tensor $c^{a}_{\;b}$ breaks SU(2)$_{PG}$ down
to U(1)
which is just the symmetry to be gauged.
Choosing the frame where the only non-zero entries of $c^a_{\; b}$ are
$c^1_{\;1} =
-c^2_{\;2} = -i$,
we recover the transformation law \p{tran4-Phi} and the constraints
\p{cons-Ph-g}.
It is easy to see that
\be
\bar {\mathcal{Z}}^+ {\mathcal{Z}}^+ = -\frac{i}{2}\,q^{+a}\,c_{ab}\,q^{+
b}\,.
\ee

In \cite{DI} (following \cite{GIOS}) an invertible equivalence
redefinition of $q^{+
a} \Rightarrow (\omega, l^{++})$
has been exploited, such that the ${\rm U}(1)$ gauge transformation in
\p{qc} is
realized as
\be
\delta \omega = -2\lambda\,, \quad \delta l^{++} = 0
\ee
(the precise form of this equivalence transformation is given in eq.
(4.26) in
\cite{DI}; it is a superfield analog
of the standard polar decomposition of a vector). Then one can fully fix
the ${\rm
U}(1)$ gauge freedom by imposing the manifestly
${\cal N}{=}4$ supersymmetric gauge
\be
\omega = 0\,.
\ee
In this gauge, the harmonic constraint in \p{qc} implies
\bea
&& \mbox{(a)} \;q^{+a}\,c_{ab}\,q^{+ b} = 4(c^{++} + l^{++})\,, \quad
\mbox{(b)} \;
V^{++} = \frac{l^{++}}{(1 + \sqrt{1 + c^{--}l^{++}})\sqrt{1 +
c^{--}l^{++}}}\,,
\nonumber \\
&& \mbox{(c)} \; D^{++}(c^{++} + l^{++}) = D^{++}l^{++} =0\,, \lb{gc2}
\eea
where $c^{\pm\pm} = c^{(ab)}u^\pm_a u^\pm_b$. After substituting the
expressions
(\ref{gc2}a) and (\ref{gc2}b) into
\p{4N-WZ} and \p{4N-FI}, the total superfield action \p{4N-gau} takes the
form:
\be
S = -{\textstyle\frac{1}{2}}\int \mu_H \, \mathscr{X}^{\,2} - i \int
\mu_A^{(-2)}
\left[\mathcal{V}\,(c^{++} + l^{++})
- \frac{c}{2}\, \frac{l^{++}}{(1 + \sqrt{1 + c^{--}l^{++}})\sqrt{1 +
c^{--}l^{++}}}\right]. \lb{Sman}
\ee

The superfield $l^{++}$ with the constraint (\ref{gc2}c)
accommodates an off-shell ${\cal N}{=}4$ multiplet (${\bf 3,4,1}$)
\cite{IL}. So, the action \p{Sman} describes a system of two
interacting off-shell ${\cal N}{=}4, d{=}1$ multiplets: (${\bf
1,4,3}$) represented by the superfield $\mathscr{X}$ and (${\bf
3,4,1}$) represented by the analytic superfield $l^{++}$. This is
the off-shell content of our ${\rm OSp}(4|2)$ model. As distinct
from the superconformal mechanics based on a single $({\bf 3, 4,
1}$) multiplet the action of which is a sum of the sigma-model type
term and WZ term of $l^{++}$ \cite{IKLech,IL}, the action \p{Sman} involves
only conformal superfield WZ term of this multiplet (the last term
in the square brackets). The interaction with the multiplet (${\bf
1,4,3}$) is accomplished through a superconformal bilinear coupling
of both multiplets (the first term in the square
brackets).\footnote{The existence of such a coupling and its
potential implications in the models of superconformal ${\cal
N}{=}4$ mechanics were noted for the first time in \cite{DI1}.}
Notice that, due to the absence of the kinetic term for $l^{++}$ in
\p{Sman}, the on-shell content of the model appears to be
drastically different from the off-shell one: the eventual component
action contains only three bosonic fields and four fermionic fields,
which are joined into some new on-shell (${\bf 3,4,1}$) multiplet
(see the next section).

\setcounter{equation}0

\section{Component actions}
\subsection{Action for {\bf (1,4,3)} supermultiplet}

The solution of the constraint~(\ref{cons-X-g-V}), (\ref{cons-X-g}) is as
follows
(in the analytic basis):
\begin{eqnarray}
\mathscr{X}&=& x +\theta^- \psi^+ + \bar\theta^- \bar\psi^+ - \theta^+
\psi^- -
\bar\theta^+ \bar\psi^-
  +\theta^-\bar\theta^- N^{++} + \theta^+\bar\theta^+ N^{--} +
(\theta^-\bar\theta^+ +
\theta^+\bar\theta^-) N \nonumber \\
&& + \,\theta^-\theta^+\bar\theta^- \Omega^+ + \bar\theta^-\bar\theta^+\theta^-
\bar\Omega^+
+ \theta^-\bar\theta^-\theta^+\bar\theta^+ D\,, \label{X-WZ}
\end{eqnarray}
where
\begin{equation}\label{N-WZ}
N^{\pm\pm} = N^{ik}u_i^\pm u_k^\pm \,,\qquad  N = i \dot x - N^{ik}u_i^+
u_k^- \,,
\qquad D = 2\ddot{x} +2i \dot{N}{}^{ik}u_i^+ u_k^- \,,
\end{equation}
\begin{equation}\label{Psi-WZ}
\psi^{\pm} = \psi^{i}u_i^\pm  \,,\qquad \bar\psi{}^{\pm} =
\bar\psi{}^{i}u_i^\pm \,,
\qquad
\Omega^{+} = 2i\dot{\psi}{}^+  \,,\qquad \bar\Omega^{+} =
-2i\dot{\bar\psi}{}^+
\end{equation}
and $x(t_A)$, $N^{ik}= N^{(ik)}(t_A)$, $\psi^{i}(t_A)$,
$\bar\psi_{i}(t_A)=(\overline{\psi^{i}})$ are $d{=}1$ fields.

Inserting~(\ref{X-WZ}) in~(\ref{4N-X}) and integrating there over the $\theta$-  and
harmonic variables \footnote{Here the harmonics
integrals $ \int du \,u^{+i}u^-_k=\frac{1}{2}\,\delta^i_k $, $ \int du
\,u^{+(i_1}u^{+i_2)}u^-_{(k_1} u^-_{k_2)}= -2\int du
\,u^{+(i_1}u^{-i_2)}u^+_{(k_1}
u^-_{k_2)}= \frac{1}{3}\,\delta^{(i_1}_{(k_1}\delta^{i_2)}_{k_2)} $ are
used.}, we
obtain
\begin{equation}\label{4N-X-WZ}
S_{\mathscr{X}} = \int dt \,\left[ \dot x\dot x -i \left(\bar\psi_k
\dot\psi^k
-\dot{\bar\psi}_k \psi^k \right) -{\textstyle\frac{1}{2}}N^{ik}N_{ik}\,
\right]\, .
\end{equation}

In the central basis the $\theta$ expansion~(\ref{X-WZ}) takes the form:
\begin{equation}  \label{sing-X0-WZ}
\mathscr{X}(t,\theta_i,\bar\theta^i)= x + \theta_i\psi^i +
\bar\psi_i\bar\theta^i +
\theta^i\bar\theta^k
N_{ik}+{\textstyle\frac{i}{2}}(\theta)^2\dot{\psi}_i\bar\theta^i
+{\textstyle\frac{i}{2}}(\bar\theta)^2\theta_i\dot{\bar\psi}{}^i +
{\textstyle\frac{1}{4}}(\theta)^2(\bar\theta)^2 \ddot{x}
\end{equation}
where $(\theta)^2\equiv \theta_i\theta^i=-2\theta^+\theta^-$,
$(\bar\theta)^2\equiv
\bar\theta^i\bar\theta_i=2\bar\theta^+\bar\theta^-\,$.
Then, from~(\ref{X0-V0}) we can identify the fields appearing in the WZ
gauge for $\mathcal{V}$ with the fields in \p{sing-X0-WZ}
\begin{equation}  \label{V0-WZ}
\mathcal{V} (t_A, \theta^+, \bar\theta^+, u^\pm) =x(t_A)- 2\,\theta^+
\psi^{i}(t_A)u^-_i  -
2\,\bar\theta^+ \bar\psi^{i}(t_A)u^-_i + 3\,\theta^+ \bar\theta^+
N^{ik}(t_A)u^-_i
u^-_k
\,.
\end{equation}
This expansion will be used to express the action~(\ref{4N-WZ}) in terms
of the
component fields.

\subsection{FI and WZ actions}

Using the ${\rm U}(1)$ gauge freedom~\p{tran4-V}, (\ref{tran4-Phi}) we can
choose WZ
gauge
\begin{equation}  \label{WZ-4N}
V^{++} =-2i\,\theta^{+}
    \bar\theta^{+}A(t_A)\,.
\end{equation}
Then
\begin{equation}  \label{4N-FI-WZ}
S_{FI} = c \int dt \,A\,.
\end{equation}

The solution of the constraint~(\ref{cons-Ph-g}) in WZ gauge~(\ref{WZ-4N}) is
\begin{equation}  \label{Ph-WZ}
{\mathcal{Z}}^+ = z^{i}u_i^+ + \theta^+ \varphi + \bar\theta^+ \phi + 2i\,
\theta^+
\bar\theta^+\nabla_{t_A}z^{i}u_i^-
\,,\qquad \bar{\mathcal{Z}}{}^+ = \bar z_{i}u^{+i} + \theta^+ \bar\phi -
\bar\theta^+
\bar\varphi + 2i\, \theta^+ \bar\theta^+ \nabla_{t_A}\bar z_{i}u^{-i}
\nonumber
\end{equation}
where
\begin{equation}\label{cov-Z}
\nabla z^k=\dot z^k + i A \, z^k\,, \qquad\nabla \bar z_k=\dot{\bar z}_k
-i A\,\bar
z_k\,.
\end{equation}
In~(\ref{Ph-WZ}), $z^{i}(t_A)$ and $\varphi(t_A)$, $\phi(t_A)$  are
$d{=}1$ fields,
bosonic and fermionic,
respectively. The fields $z^i$ form a complex doublet of the R-symmetry ${\rm
SU}(2)$ group, while the fermionic fields
are singlets of the latter. Another (``mirror'') R-symmetry ${\rm SU}(2)$
is not
manifest in the present approach: the bosonic
fields are its singlets, while the fermionic fields form a doublet with
respect to it.

Inserting expressions~(\ref{Ph-WZ}) and (\ref{V0-WZ}) in the
action~(\ref{4N-WZ}) and integrating over $\theta\,$s and harmonics, we
obtain a component form of the WZ action
\begin{eqnarray}\label{4N-WZ-WZ}
S_{WZ} &=& {\textstyle\frac{i}{2}}\int dt \,\left(\bar z_k \nabla z^k -
\nabla \bar z_k
\, z^k \right)x - {\textstyle\frac{1}{2}}\int dt \,N^{ik}\bar z_i z_k
\\
\nonumber && \, + {\textstyle\frac{1}{2}}\int dt \, \Big[\psi^k
\left(\bar\varphi\,
z_k+
\bar z_k\phi\right) + \bar\psi^k \left(\bar{\phi\,\,}\! z_k- \bar
z_k\varphi\right) - x
\left(\bar{\phi\,\,}\! \phi+ \bar\varphi\,\varphi\right)\Big] \, .
\end{eqnarray}
The fermionic fields $\phi, \varphi$ are auxiliary. The action is
invariant under
the residual local ${\rm U}(1)$ transformations
\be
A' = A - \dot{\lambda}_0\,, \quad z^i{}' = e^{i\lambda_0}z^i\,, \;
\bar{z}_i{}' =
e^{-i\lambda_0}\bar{z}_i \label{res-ga}
\ee
(and similar phase transformations of the fermionic fields).

The total component action is a sum of \p{4N-X-WZ}, \p{4N-FI-WZ} and
\p{4N-WZ-WZ}.
Eliminating the auxiliary fields $N^{ik}$, $\phi$,  $\bar\phi$, $\varphi$,
$\bar\varphi$,
from this sum by their algebraic equations of motion,
\begin{eqnarray}
&& N_{ik} =- {\textstyle\frac{1}{2}}\,z_{(i} \bar z_{k)}\,,
\label{4N-eq-N} \\
&& \phi =-\frac{\bar\psi^{k} z_{k}}{x} \,, \quad \bar\phi =\frac{\psi^{k}
\bar
z_{k}}{x} \,,
\quad \varphi =-\frac{\psi^{k} z_{k}}{x}\,, \quad \bar\varphi
=-\frac{\bar\psi^{k} \bar
z_{k}}{x}\,,  \label{4N-eq-phi}
\end{eqnarray}
and making the redefinition
\begin{equation}\label{4N-nZ}
z^\prime{}^i =  x^{1/2}\,z^i\,,
\end{equation}
we obtain that the action~(\ref{4N-gau}) in WZ gauge takes the following
on-shell
form (we omitted the primes
on $z$)
\begin{eqnarray}
S &=& S_b+ S_f\,, \label{4N-ph}\\
S_b &=&  \int dt \,\Big[\dot x\dot x  + {\textstyle\frac{i}{2}} \left(\bar
z_k \dot
z^k -
\dot{\bar z}_k z^k\right)-\frac{(\bar z_k
z^{k})^2}{16x^2} -A \left(\bar z_k z^{k} -c \right) \Big] \,,\label{bose}\\
S_f &=&  -i\, \int dt \,\left( \bar\psi_k \dot\psi^k -\dot{\bar\psi}_k
\psi^k \right)-
    \int dt \, \frac{\psi^{i}\bar\psi^{k} z_{(i} \bar z_{k)}}{x^2}\,.
\label{fermi}
\end{eqnarray}

It is still invariant under the gauge transformations \p{res-ga}. The $d{=}1$
connection
$A(t)$ in \p{bose} is the Lagrange multiplier for the constraint
\begin{equation}\label{con}
\bar z_k z^{k}=c\,.
\end{equation}
After varying with respect to $A$, the action \p{4N-ph} is gauge invariant
only with
taking into account
this algebraic constraint which is gauge invariant by itself. It is
convenient to
fully fix the residual
gauge freedom by choosing the phases of $z^1$ and $z^2$ opposite to
each other.
In this gauge,
the constraint \p{con} is solved by
\begin{equation}\label{ga-u1}
z^{1} = \kappa \cos{\textstyle\frac{\gamma}{2}}\,e^{i\alpha/2}\,,\qquad
z^{2}= \kappa \sin{\textstyle\frac{\gamma}{2}}\,e^{-i\alpha/2}\,,\qquad
\kappa^2=c\,.
\end{equation}
In terms of the newly introduced  fields the action~(\ref{4N-ph}) takes
the form
\begin{eqnarray}
S &=& S_b+ S_f\,, \label{4N-ph1}\\
S_b &=&  \int dt \,\Big[\dot x\dot x  -\frac{c^2}{16\,x^2} - \frac{c}{2}\,
\cos\gamma\, \dot \alpha  \Big] \,,\label{bose1}\\
S_f &=&  -i\, \int dt \,\left( \bar\psi_k \dot\psi^k -\dot{\bar\psi}_k \psi^k
\right) \nonumber \\
&& +\, \frac{c}{2} \int dt \, \frac{ \cos\gamma \left( \psi^1 \bar\psi_1
+\psi^2
\bar\psi_2 \right)- \sin\gamma \left( e^{i\alpha}\psi^2 \bar\psi_1
+e^{-i\alpha}\psi^1
\bar\psi_2 \right) }{x^2}\,.\label{fermi1}
\end{eqnarray}

Unconstrained fields in the action~(\ref{4N-ph1}), three bosons $x$,
$\gamma$,
$\alpha$ and
four fermions $\psi^k$, $\bar\psi_k$, constitute some on-shell ({\bf 3,4,1})
supermultiplet.
As opposed to the ({\bf 3,4,1}) supermultiplet considered
in~\cite{IKLech,IKL,IKLecht}
the action~(\ref{bose1}) contains ``true'' kinetic term only for one
bosonic component
which also possesses the conformal potential, whereas two other fields
parametrizing
the coset SU(2)$_R$/U(1)$_R$
are described by a WZ term and so become a sort of ispospin degrees of
freedom (target SU(2) harmonics) upon
quantization.
The realization of ${\rm{OSp}}(4|2)$ superconformal transformations on
these fields
will be given in the next section.

It should be stressed that the considered model realizes a new mechanism of
generating conformal potential $\sim 1/x^2$ for
the field $x(t)$. Before eliminating auxiliary fields, the
component
action contains no explicit term
of this kind. It arises as a result of varying with respect to
the Lagrange
multiplier $A(t)$ and making use of the arising constraint \p{con}. As we shall see
soon, in quantum theory
this new
mechanism entails a quantization
of the constant $c\,$. In the ${\rm SU}(1,1|2)$ superconformal quantum
mechanics,
the strength of the conformal
potential appears in the $su(1,1|2)$ algebra as a constant central charge
\cite{IKL,Wyl,GalLe}.
In our model such an option does not exist
since the superalgebra $osp(4|2)$ does not alow a central extension.

Notice that an equivalent component action can be obtained starting from the
superfield action \p{Sman} which
corresponds to another choice of the gauge with respect to U(1)
transformations. As
distinct from the WZ gauge
used in this section, the gauge corresponding to \p{Sman} preserves the
manifest
${\cal N}{=}4$ supersymmetry
and does not exhibit any residual gauge freedom. The component bosonic
sector of
\p{Sman} involves
one physical $x = \mathscr{X}|_{\theta =0}$ and three bosonic fields
$y^{(ik)}$ from
the $({\bf 3,4,1})$ superfield
$l^{++}$. They form a 3-vector with respect to ${\rm SU}(2)_R$ ($l^{++} +
c^{++} =
y^{(ik)}u^+_iu^+ + \mbox{$\theta$-dependent terms}$).
By an algebraic constraint, with the auxiliary field of $({\bf 3,4,1})$ as a
Lagrange multiplier, the fields
$y^{(ik)}$ are confined  to parametrize a sphere $S^2$. This constraint
plays a role
analogous to \p{con}. The gauge-invariant
fields $y^{(ik)}$ are related to the doublet fields $z^i, \bar z_k$ via the
well-known first Hopf map (see also sect. below).
The relation (\ref{gc2}a) is in fact a superfield version of this map.
Thus, one
again ends up with 3 bosonic fields
and 4 fermionic fields forming an irreducible on-shell multiplet.

It is also worth noting that this reduction of two independent off-shell
${\cal
N}{=}4$ multiplets ({\bf 3,4,1}) and ({\bf 1,4,3})
to a smaller on-shell ${\cal N}{=}4$ multiplet somewhat resembles the
procedure of
ref. \cite{KLSh} in which some irreducible
${\cal N}{=}4$ multiplets with four physical fermions are generated from
pairs of
other multiplets of this type by
identifying fermionic fields in the multiplets forming the pair. In our
case such
identification arises as one of the
algebraic equations of motion, eq. \p{4N-eq-phi}. In this connection, it
would be
interesting to inquire whether the component action
\p{4N-ph1} can be independently re-derived from an alternative (dual)
superfield
action corresponding to
some nonlinear version of the off-shell multiplet ({\bf 3,4,1}).

\subsection{${\cal N}{=}4$ superconformal symmetry in WZ gauge}

The transformations and their generators look most transparent
in terms of the SU(2) doublet quantities $z^{k}$ and $\bar z_{k}$.

To determine the superconformal transformations of component fields, we
should know the
appropriate compensating gauge transformations needed to preserve the WZ
gauge~(\ref{WZ-4N}).
For supertranslations and superconformal boosts the parameter of the compensating
gauge transformations is as follows
\begin{equation}\label{tran-g-WZ}
\lambda=2i\left[ (\theta^+\bar\varepsilon^- -\bar\theta^+\varepsilon^-) + t_A\,
(\theta^+\bar\eta^- -\bar\theta^+\eta^-) \right]A
\end{equation}
where
\be
\varepsilon^- := \varepsilon^iu^-_i\,, \quad \eta^- := \eta^iu^-_i\,.
\ee
Taking this into account, we obtain the relevant infinitesimal OSp(4$|$2)
transformations:
\begin{equation}\label{2str-X}
\delta x=-\omega_i\psi^i+ \bar\omega^i\bar\psi_i,
\end{equation}
\begin{equation}\label{2str-Psi}
\delta \psi^i=\frac{\bar\omega_k z^{(i} \bar z^{k)}}{2x}-i
\bar\omega^i\dot x +i\bar\eta^i x,
\end{equation}
\begin{equation}\label{2str-Z}
\delta z^i=\frac{\omega^{(i}\psi^{k)} +\bar\omega^{(i}\bar\psi^{k)}}{x}\,z_k\,,
\end{equation}
\begin{equation}\label{2str-A}
\delta A=0\,,
\end{equation}
where $\omega^i = \varepsilon^i + t\, \eta^i\, $.

Now, using the N\"other procedure,
we can directly find the classical generators
of the supertranslations
\begin{equation}\label{Q-cl}
{Q}^i =p\, \psi^i- i\frac{z^{(i} \bar z^{k)}\psi_k}{x}\, ,\qquad
\bar{Q}_i=p\,
\bar\psi_i+
i\frac{z_{(i} \bar z_{k)}\bar\psi^k}{x}\,
\end{equation}
where $ p\equiv 2\dot x$, as well as of the superconformal boosts:
\begin{equation}\label{S-cl}
{S}^i =-2\,x \psi^i + t\,{Q}^i,\qquad \bar{S}_i=-2\,x
\bar\psi_i+t\,\bar{Q}_i\,.
\end{equation}
The remaining (even) generators of the supergroup ${\rm{OSp}}(4|2)$ can be
found by
evaluating anticommutators
of the above odd generators among themselves.

As follows from the action~(\ref{4N-ph}), the ${\rm{SU}}(2)$ spinor
variables are
canonically
self--conjugate due to the presence of second-class constraints for their
momenta.
As a result, non-vanishing canonical Dirac brackets (at equal times) have the
following form
\begin{equation}\label{CDB}
[x, p]_{{}_D}= 1, \qquad [z^i, \bar z_j]_{{}_D}= -i\delta^i_j, \qquad
\{\psi^{ii^\prime},
\psi^{kk^\prime}\}_{{}_D}= {\textstyle\frac{i}{2}}\,\epsilon^{i
k}\epsilon^{i^\prime
k^\prime}
\end{equation}
where we introduced the notations
\begin{equation}\label{psi-22}
\psi^{ii^\prime}=(\psi^{i1^\prime},\psi^{i2^\prime})=(\psi^i,
\bar\psi^i),\qquad
(\overline{\psi^{ii^\prime}})=\psi_{ii^\prime}=\epsilon_{i
k}\epsilon_{i^\prime
k^\prime}
\psi^{kk^\prime}\,, \quad (\epsilon_{12} = \epsilon^{21} = 1).
\end{equation}

Using Dirac brackets~(\ref{CDB}), we arrive at the following closed
superalgebra:
\begin{eqnarray}
&& \{{Q}^{ai^\prime i}, {Q}^{bk^\prime k}\}_{{}_D}=
2i\left(\epsilon^{ik}\epsilon^{i^\prime
k^\prime} T^{ab}+\alpha \epsilon^{ab}\epsilon^{i^\prime k^\prime}
J^{ik}-(1+\alpha)
\epsilon^{ab}\epsilon^{ik} I^{i^\prime k^\prime}\right)\,,\label{DB-Q-g} \\
&& [T^{ab}, T^{cd}]_{{}_D}= -\epsilon^{ac}T^{bd}
-\epsilon^{bd}T^{ac},\quad [J^{ij},
J^{kl}]_{{}_D}= -\epsilon^{ik}J^{jl} -\epsilon^{jl}J^{ik},\nonumber \\
&& [I^{i^\prime
j^\prime}, I^{k^\prime l^\prime}]_{{}_D}= -\epsilon^{ik}I^{j^\prime l^\prime}
-\epsilon^{j^\prime
l^\prime}I^{i^\prime k^\prime},\label{DB-J} \\
&& [T^{ab}, {Q}^{ci^\prime i}]_{{}_D}=\epsilon^{c(a}{Q}^{b)i^\prime i}
,\quad [J^{ij},
{Q}^{ai^\prime k}]_{{}_D}=\epsilon^{k(i}{Q}^{ai^\prime j)},\quad
[J^{i^\prime
j^\prime},
{Q}^{ak^\prime i}]_{{}_D}=\epsilon^{k^\prime (i^\prime}{Q}^{aj^\prime )
i}\label{DB-JQ}
\end{eqnarray}
where $\alpha=-\frac{1}{2}$. In~(\ref{DB-Q-g})-(\ref{DB-JQ}) we use the
notation
\begin{equation}\label{not-Q}
{Q}^{21^\prime i}=-{Q}^{i}\,,\quad {Q}^{22^\prime i}=-\bar{Q}^{i}\,,
\qquad\qquad
{Q}^{11^\prime i}={S}^{i}\,,\quad {Q}^{12^\prime i}=\bar{S}^{i}\,,
\end{equation}
\begin{equation}\label{not-T}
T^{22}=H\,,\quad T^{11}=K\,,\quad T^{12}=-D\,.
\end{equation}
The explicit expressions for the generators are
\begin{eqnarray}\label{H-cl}
H &=&{\textstyle\frac{1}{4}}\,p^2  +\frac{(\bar z_k z^{k} )^2}{16\,x^2} +
    \frac{z^{i} \bar z^{j} \,\psi_{i}{}^{k^\prime}
\psi_{jk^\prime}}{2\,x^2}\,,
\\ \label{K-cl}
K &=&x^2  -t\,x p +
    t^2\, H\,,
\\ \label{D-cl}
D &=&-{\textstyle\frac{1}{2}}\,x p +
    t\, H\,,
\\ \label{T-cl}
J^{ij} &=& i\left[ z^{(i} \bar z^{j)}+
\psi^{ik^\prime}\psi^{j}{}_{k^\prime}\right]\,,
\\ \label{I-cl}
I^{i^\prime j^\prime} &=& i \psi^{ki^\prime}\psi_{k}{}^{j^\prime}\,.
\end{eqnarray}
The relations~(\ref{DB-Q-g})-(\ref{DB-JQ}) provide the standard form of
the superalgebra
$D(2,1;-\frac12)\simeq {\rm{OSp}}(4|2)$ (see,
e.g.,~\cite{Sorba,BILS,IKLech}). Bosonic
generators $T^{ab}=T^{ba}$, $J^{ik}=J^{ki}$, $I^{i^\prime
k^\prime}=I^{k^\prime
i^\prime}$
form mutually commuting $su(1,1)$, $su(2)$ and $su{\,}^\prime(2)$ algebras,
respectively.

The expression~(\ref{H-cl}) is precisely the canonical Hamiltonian
obtained from the
action~(\ref{4N-ph}).
Owing to the $A$-term in~(\ref{4N-ph}), there is also the first-class
constraint
\begin{equation}\label{D0-con}
D^0\equiv \bar z_k z^{k} -c \approx 0\,,
\end{equation}
which should be imposed on wave functions in quantum case.

In the next section we shall construct a quantum realization of
$D(2,1;-\frac12)$
superalgebra
given above.

\setcounter{equation}0
\section{${\rm{OSp}}(4|2)$ quantum mechanics}

\subsection{Bosonic limit and fuzzy sphere}
In order to understand the specific features of our model better, we begin by
quantizing it
in the bosonic limit, with all fermionic fields discarded.
It reveals an interesting deviation from the standard conformal quantum
mechanics of
de\,Alfaro, Fubini and Furlan~\cite{AFF}:
besides the standard dilatonic variable $x(t)$ with the conformal
potential, it also
contains a fuzzy sphere~\cite{Mad,IMT,Has} represented
by the SU(2) spinor variables $z^i(t), \bar z{}^i(t)$. As a result, the
relevant wave
functions are non-trivial SU(2) multiplets, as
opposed to the singlet wave function of the standard conformal mechanics. The
strength of the conformal potential proves to coincide with the
eigenvalue of the ${\rm SU}(2)$ Casimir operator (i.e. ``spin'') and so is
quantized.

The pure bosonic model is described by the action~(\ref{bose}). The
corresponding
canonical
Hamiltonian reads
\begin{equation}\label{bose-H0}
H_0 =  \frac{1}{4} \left[ p^2 + \frac{(\bar z_k
z^{k})^2}{4x^2} \right] +A \left(\bar z_k z^{k} -c  \right).
\end{equation}
Here $p=2\dot x$ is the canonical momentum for the coordinate $x$. Canonical
momentum for the field $A$ is vanishing, $p_A=0$.
This constraint and the fact that the field $A$ appears in the
action~(\ref{bose})
linearly, suggest to treat $A$ as the Lagrange multiplier
for the constraint
\begin{equation}\label{D0-con1}
D^0-c\equiv \bar z_k z^{k} -c \approx 0\,.
\end{equation}
Expressions for the canonical momenta $p_i$ and $\bar p^i$ for the
$z$-variables,
$[z^i, p_j]_{{}_P}= \delta^i_j$, $[\bar z_i, p^j]_{{}_P}= \delta^j_i$, are
the
second-class
constraints
\begin{equation}\label{G-const}
G_k\equiv p_k-{\textstyle\frac{i}{2}}\,\bar z_k\approx 0\,,\qquad
\bar G^k\equiv \bar p^k+{\textstyle\frac{i}{2}}\, z^k\approx 0\,,\qquad
[G_k, \bar G^l]_{{}_P}=  -i\delta^l_k.
\end{equation}
Using Dirac brackets for them
$$
[A, B]_{{}_D}=[A, B]_{{}_P}+i[A, G_k]_{{}_P}[\bar G^k, B]_{{}_P}-i[A, \bar
G^k]_{{}_P}[G_k, B]_{{}_P}
$$
we eliminate the spinor momenta $p_i$ and $\bar p^i$.
Dirac brackets for the residual variables are
\begin{equation}\label{DB}
[x, p]_{{}_D}= 1, \qquad [z^i, \bar z_j]_{{}_D}= -i\delta^i_j.
\end{equation}

To finish with the classical description, we point out that the spinor variables
describe a two--sphere.
Namely, using the first Hopf map we introduce three U(1) gauge invariant variables
\begin{equation}\label{cl-y}
y_a= {\textstyle\frac{1}{2}}\,\bar z_i(\sigma_a)^i{}_j z^j
\end{equation}
where $\sigma_a$, $a=1,2,3$ are Pauli matrices. The
constraint~(\ref{D0-con1})
suggests that these variables
parameterize a two--sphere with the radius $c/2$:
\begin{equation}\label{cl-sp}
y_a y_a=(z^{k}\bar z_k)^2/4\approx c^2/4\,.
\end{equation}
The group of motion of this 2-sphere is of course the R-symmetry ${\rm SU}(2)$ group
acting
on the doublet indices $i, k$ and triplet indices $a$. In terms of the new
variables~(\ref{cl-y}) the
Hamiltonian~(\ref{bose-H0}), up to
terms vanishing on the constraints, takes the form
\begin{equation}\label{bose-Hy}
H =  \frac{1}{4} \left[ p^2 + \frac{y_a y_a}{x^2} \right] \,.
\end{equation}
It is worth pointing out that \p{cl-y} is none other than the WZ gauge
counterpart
of the superfield Hopf map (\ref{gc2}a).

At the quantum level, the algebra of the canonical operators obtained from
the
algebra of Dirac brackets is
(quantum operators are denoted by the appropriate capital letters),
\begin{equation}\label{bose-cB}
[X, P] = i\,, \qquad [Z^i, \bar Z_j] = \delta^i_j \,.
\end{equation}
Then it is easy to check that the quantum counterparts of the
variables~(\ref{cl-y})
\begin{equation}\label{qu-y}
Y_a= {\textstyle\frac{1}{2}}\,\bar Z_i(\sigma_a)^i{}_j Z^j
\end{equation}
form the SU(2) algebra
\begin{equation}\label{Y-cB}
[Y_a, Y_b] = i\,\epsilon_{abc} Y_c\,.
\end{equation}
Notice that no ordering ambiguity is present in the definition \p{qu-y}.

Moreover, the direct calculation yields
\begin{equation}\label{Y2}
Y_a Y_a = {\textstyle\frac{1}{2}}\,\bar Z_k Z^{k}
\left({\textstyle\frac{1}{2}}\,\bar
Z_k Z^{k}+1\right)
\end{equation}
and, due to the constraints (for definiteness, we adopt $\bar Z_k
Z^{k}$--ordering
in it), one gets
\begin{equation}\label{Y2-1}
Y_a Y_a = \frac{c}{2} \left(\frac{c}{2}+1\right)\,.
\end{equation}
But the relations~(\ref{Y-cB}) and~(\ref{Y2-1}) are the definition of the
{\it fuzzy
sphere} coordinates~\cite{Mad}.
Thus the angular variables, described, at the
classical level, by spinor variables $z^i$ or vector variables $y_a$,
after quantization acquire a nice interpretation of the fuzzy sphere
coordinates.
Comparing the expressions~(\ref{Y2}) and~(\ref{Y2-1}), we observe that
upon quantization the radius of the sphere changes as  $\frac{c^2}{4}\to
\frac{c}{2}
\left(\frac{c}{2}+1\right)\,$.

As suggested by the relation~(\ref{Y-cB}), the fuzzy sphere coordinates $Y_a$ are the
generators
of $su(2)_R$ algebra and the relation~(\ref{Y2-1}) fixes the value of its
Casimir
operator, with $c$ being the relevant
${\rm SU}(2)$ spin (``fuzzyness''). Then it follows that $c$ is quantized,
$c \in
\mathbb{Z}$.
Actually, from the standpoint of the supergroup ${\rm{OSp}}(4|2)$, this
$su(2)$
algebra is
just a quantum version of the $su(2)$ generated by generators $J^{ik}$
defined in
\p{T-cl}.

The wave functions inherit this internal symmetry through a  dependence on
additional ${\rm SU}(2)$ spinor degrees of freedom. Let us consider the
following
realization
for the operators $Z^i$ and $\bar Z_i$
\begin{equation}\label{bo-re-Z}
\bar Z_i=v^+_i, \qquad Z^i=  \partial/\partial v^+_i
\end{equation}
where $v^+_i$ is a commuting complex SU(2) spinor.
Then the constraint~(\ref{D0-con1}) on wave function $\Phi(x,v^+_i)$
\begin{equation}\label{q-con1}
D^0 \Phi=\bar Z_i Z^i \Phi=v^+_i\frac{\partial}{\partial v^+_i}\,\Phi=c\,\Phi
\end{equation}
leads to the polynomial dependence of it on $v^+_i$:
\begin{equation}\label{wf-rep}
\Phi(x,v^+_i) = \phi_{k_1\ldots k_{c}}(x)v^{+k_1}\ldots v^{+k_{c}} \,.
\end{equation}
Thus, as opposed to the model of ref.~\cite{AFF}, in our case the
$x$--dependent
wave function
carries an irreducible spin $c/2$ representation of the group SU(2), being
an SU(2)
spinor of the rank $c$.

Using~(\ref{bose-Hy}) and~(\ref{Y2-1}) we see that on physical states the
quantum
Hamiltonian is
\begin{equation}\label{H-qu-bo}
\mathbf{H} =\frac{1}{4}\,\left(P^2  +\frac{g}{X^2} \right)\,,
\end{equation}
where
\begin{equation}\label{bose-g}
g = \frac{c}{2} \left(\frac{c}{2}+1\right)\,.
\end{equation}
It is easy to show that the SU(1,1) Casimir operator takes the value
${\textstyle\frac{1}{4}}\, g - {\textstyle\frac{3}{16}}$ (for further
details, see
the next sections).
Thus, like in~\cite{AFF}, on the fields $\phi_{k_1\ldots k_{c}}(x)$
the unitary irreducible representations of the group SU(1,1) are realized,
despite
the fact that the wave function
is now multi-component, with $(c + 1)$ independent components.
Requiring the wave function $\Phi(v^+)$ to be single--valued once again
leads to the
condition that $c\in \mathbb{Z}$.
This quantization of parameter $c$ could be important for the possible black hole
interpretation of the considered variant of conformal mechanics.

Note that the new variables $v^+_i$ can be treated as a half of the target space
harmonic-like
variables $v^+_i, v^-_i$ (though without the familiar constraint $v^{+ i}v_i^- \sim
const$). The harmonic
interpretation could be made more literal using a different, mixed Dirac-
Gupta-Bleuler quantization for $z$ variables
along the line of ref. \cite{ABS}.

\subsection{Operator realization of ${\rm{OSp}}(4|2)$ superalgebra}

Here we extend the bosonic-limit consideration to the whole OSp$(4|2)$
mechanics.

Quantum operators of physical coordinates and momenta satisfy the quantum
brackets,
obtained in the standard way from~(\ref{CDB}) (by multiplying the latter
by $i$):
\begin{equation}\label{cB}
[X, P] = i\,, \qquad [Z^i, \bar Z_j] = \delta^i_j \,, \qquad \{\Psi^i,
\bar\Psi_j\}=
-{\textstyle\frac{1}{2}}\,\delta^i_j \,.
\end{equation}

Quantum supertranslation and superconforml boost generators are uniquely
defined by the classical expressions (\ref{Q-cl}), (\ref{S-cl}). They
appear to be
linear in the odd operators:
\begin{equation}\label{Q-qu}
\mathbf{Q}^i =P \Psi^i- i\frac{Z^{(i} \bar Z^{k)}\Psi_k}{X}\, ,\qquad
\bar\mathbf{Q}_i=P
\bar\Psi_i+ i\frac{Z_{(i} \bar Z_{k)}\bar\Psi^k}{X}\,,
\end{equation}
\begin{equation}\label{S-qu}
\mathbf{S}^i =-2\,X \Psi^i + t\,\mathbf{Q}^i,\qquad \bar\mathbf{S}_i=-2\,X
\bar\Psi_i+ t\,\bar\mathbf{Q}_i\,.
\end{equation}

Evaluating the anticommutators of the odd generators (\ref{Q-qu}),
(\ref{S-qu}), one
determines
uniquely the full set of quantum generators of superconformal algebra
$D(2,1;-\frac12)$.
We obtain
\begin{eqnarray}\label{H-qu}
\mathbf{H} &=&{\textstyle\frac{1}{4}}\,P^2  +\frac{(\bar Z_k
Z^{k})^2+2\bar Z_k
Z^{k}}{16X^2} +
    \frac{Z^{(i} \bar Z^{k)} \Psi_{(i}\bar\Psi_{k)}}{X^2}\,,
\\ \label{K-qu}
\mathbf{K} &=&X^2  - t\,{\textstyle\frac{1}{2}}\,\{X, P\} +
    t^2\, \mathbf{H}\,,
\\ \label{D-qu}
\mathbf{D} &=&-{\textstyle\frac{1}{4}}\,\{X, P\} +
    t\, \mathbf{H}\,,
\\ \label{T-qu}
\mathbf{J}^{ik} &=& i\left[ Z^{(i} \bar Z^{k)}+
2\Psi^{(i}\bar\Psi^{k)}\right]\,,
\\ \label{I-qu}
\mathbf{I}^{1^\prime 1^\prime} &=& -i\Psi_k\Psi^k\,,\qquad
\mathbf{I}^{2^\prime
2^\prime} =
i\bar\Psi^k\bar\Psi_k\,,\qquad \mathbf{I}^{1^\prime 2^\prime}
=-{\textstyle\frac{i}{2}}\,
[\Psi_k,\bar\Psi^k]\,.
\end{eqnarray}
It can be directly checked that the generators~(\ref{Q-qu})--(\ref{I-qu})
indeed form
the $D(2,1;-\frac12)$ superalgebra which is obtained form the DB
superalgebra~(\ref{DB-Q-g})-(\ref{DB-JQ}) by changing altogether DB by
(anti)commutators and multiplying
the right-hand sides by $i$.

The second--order Casimir operator of $D(2,1;-\frac12)$ is given by the
following expression~\cite{Je}
\begin{equation}\label{qu-Cas}
\mathbf{C}_2=\mathbf{T}^2 -{\textstyle\frac{1}{2}}\, (\mathbf{J}^2+
\mathbf{I}^2) + {\textstyle\frac{i}{4}}\, \mathbf{Q}^{ai^\prime
i}\mathbf{Q}_{ai^\prime i}
\end{equation}
(the quantum ${\rm SU}(1,1)$ generators $\mathbf{T}^{ab}$ are defined in
terms of
the generators \p{H-qu} - \p{D-qu}
by the same formulas \p{not-T}). Using the relations
\begin{eqnarray}\label{q-Cas-3}
\mathbf{T}^2&\equiv& {\textstyle\frac{1}{2}}\,
\mathbf{T}^{ab}\mathbf{T}_{ab} =
{\textstyle\frac{1}{16}}\, \left[(\bar Z_k Z^{k})^2+2\bar Z_k Z^{k}\right]  +
Z^{(i} \bar Z^{k)}\Psi_{(i}\bar\Psi_{k)} -
{\textstyle\frac{3}{16}}\,,
\\
\label{q-Cas-2}
\mathbf{J}^2&\equiv& {\textstyle\frac{1}{2}}\,
\mathbf{J}^{ik}\mathbf{J}_{ik} =
{\textstyle\frac{1}{4}}\, \left[(\bar Z_k Z^{k})^2+2\bar Z_k Z^{k}\right] -
{\textstyle\frac{3}{2}}\,\left(\Psi_i\Psi^i\, \bar\Psi^k\bar\Psi_k -
\Psi_i \bar\Psi^i
\right) - 2 Z^{(i} \bar Z^{k)}\Psi_{(i}\bar\Psi_{k)},
\\
\label{q-Cas-1}
\mathbf{I}^2&\equiv& {\textstyle\frac{1}{2}}\, \mathbf{I}^{i^\prime k^\prime
}\mathbf{I}_{i^\prime k^\prime }
={\textstyle\frac{1}{2}}\,\{\bar\mathbf{I},\mathbf{I}\} -(\mathbf{I}_3)^2
= {\textstyle\frac{3}{2}}\,\left(\Psi_i\Psi^i\, \bar\Psi^k\bar\Psi_k - \Psi_i
\bar\Psi^i
\right)+ {\textstyle\frac{3}{4}}
\end{eqnarray}
together with
\begin{equation}\label{QQ-quCas}
{\textstyle\frac{i}{4}}\, \mathbf{Q}^{ai^\prime i}\mathbf{Q}_{ai^\prime i} =
{\textstyle\frac{i}{4}}\, [\mathbf{Q}^{i},\bar \mathbf{S}_{i}] +
{\textstyle\frac{i}{4}}\,
[\bar \mathbf{Q}_{i}, \mathbf{S}^{i}] = - 2 Z^{(i} \bar
Z^{k)}\Psi_{(i}\bar\Psi_{k)}
+{\textstyle\frac{1}{2}}\,,
\end{equation}
we find that $\mathbf{C}_2$ takes the form
\begin{equation}\label{qu-Cas-12}
\mathbf{C}_2=-{\textstyle\frac{1}{16}}\, \left[(\bar Z_k Z^{k})^2+2\bar Z_k
Z^{k}+1\right]
\,.
\end{equation}
Using~(\ref{qu-Cas-12}), we can rewrite quantum Hamiltonian (\ref{H-qu})
in the
following
equivalent suggesting form:
\begin{equation}\label{H-qu-1}
\mathbf{H} ={\textstyle\frac{1}{4}}\,P^2 -\frac{\mathbf{C}_2}{X^2}
-\frac{1}{16X^2} +
    \frac{Z^{(i} \bar Z^{k)} \Psi_{(i}\bar\Psi_{k)}}{X^2}\,.
\end{equation}

An important observation is that the following  quantities belonging to the
enveloping
algebra of $osp(4|2)$ superalgebra
\begin{eqnarray}\label{M}
\mathbf{M}&\equiv& 4\mathbf{T}^2 -(\mathbf{J}^2+ \mathbf{I}^2)
+{\textstyle\frac{3i}{4}}\,
\mathbf{Q}^{ai^\prime i}\mathbf{Q}_{ai^\prime i} \,,
\\
\label{M-ai}
\mathbf{M}^{ik\!,\,i^\prime k^\prime}&\equiv&  \{ \mathbf{J}^{ik},
\mathbf{I}^{i^\prime k^\prime}\}
+i\mathbf{Q}^{b{}^{\scriptstyle (}i^\prime (i}\mathbf{Q}_{b}{}^{k^\prime
{}^{\scriptstyle )}
k )} \,,
\\
\label{M-ik}
\mathbf{M}^{ai^\prime i}&\equiv&  {\textstyle\frac{i}{2}}\,
\{ \mathbf{T}^{a}_b , \mathbf{Q}^{bi^\prime i} \} +
{\textstyle\frac{i}{4}}\,
\{ \mathbf{J}^{i}_j, \mathbf{Q}^{ai^\prime j}\} +
{\textstyle\frac{i}{4}}\,
\{ \mathbf{I}^{i^\prime}_{j^\prime}, \mathbf{Q}^{aj^\prime i} \}
\end{eqnarray}
form a linear finite--dimensional representation of OSp(4$|$2):
$$
[\mathbf{M}, \mathbf{Q}^{ai^\prime i}]= \mathbf{M}^{ai^\prime i},\qquad
[\mathbf{M}^{ik\!,\,i^\prime k^\prime}, \mathbf{Q}^{bj^\prime
j}]=-4\epsilon^{j(i}
\epsilon^{j^\prime{}^{\scriptstyle
(}i^\prime}\mathbf{M}^{ak^\prime{}^{\scriptstyle
)} k)},
$$
$$
[\mathbf{M}^{ai^\prime i}, \mathbf{Q}^{bk^\prime
k}]=-{\textstyle\frac{i}{2}}\,
\epsilon^{ab}
\epsilon^{i^\prime k^\prime}\epsilon^{ik}\mathbf{M}
-{\textstyle\frac{i}{2}}\,
\epsilon^{ab}\mathbf{M}^{ik\!,\,i^\prime k^\prime}\,.
$$
For the particular representation of generators given by eqs.
(\ref{q-Cas-3})-(\ref{q-Cas-1})
all quantities  (\ref{M})-(\ref{M-ik}) identically vanish:
\begin{equation}\label{id-alg}
\mathbf{M}=0 \,,
\qquad
\mathbf{M}^{ik\!,\,i^\prime k^\prime}=0 \,,
\qquad
\mathbf{M}^{ai^\prime i}=0\,.
\end{equation}
As a consequence of these identities, there arises the relation
\begin{equation}\label{id-Cas}
\mathbf{T}^2 +{\textstyle\frac{1}{2}}\, (\mathbf{J}^2+ \mathbf{I}^2)
= -3\mathbf{C}_2\,.
\end{equation}
Thus, for an irreducible representation of
$D(2,1;-\frac{1}{2})$ with
a fixed $\mathbf{C}_2$ (see
(\ref{qu-Cas-ev}) below) the values of the Casimir operators $\mathbf{T}^2$,
$\mathbf{J}^2$, $\mathbf{I}^2$ of three bosonic subgroups $sl(2,R)$, $su(2)$,
$su^\prime(2)$ prove to be related as in \p{id-Cas}.

\subsection{Quantum spectrum}

The Hamiltonian (\ref{H-qu}) and the $sl(2,R)$ Casimir operator
(\ref{q-Cas-3}) can
be represented as
\begin{equation}\label{H-qu-g}
\mathbf{H} =\frac{1}{4}\,\left(P^2  +\frac{\hat g}{X^2} \right)\,,
\end{equation}
\begin{equation}\label{q-Cas-g}
\mathbf{T}^2 = {\textstyle\frac{1}{4}}\,\hat g - {\textstyle\frac{3}{16}}\,,
\end{equation}
where
\begin{equation}\label{hat-g}
\hat g \equiv {\textstyle\frac{1}{2}}\,\bar Z_k Z^{k}
\left({\textstyle\frac{1}{2}}\,\bar
Z_k Z^{k}+1\right) + 4 Z^{(i} \bar Z^{k)} \Psi_{(i}\bar\Psi_{k)}\,.
\end{equation}
The operators (\ref{H-qu-g}) and (\ref{q-Cas-g}) formally look like those 
given in the ${\rm SU}(1,1)$ model of ~\cite{AFF}.
However, there is an essential difference. Whereas the quantity $\hat g$ is a
constant in the ${\rm SU}(1,1)$ model, in our case $\hat g$ is an operator 
even in the bosonic sector taking fixed, but different, constant values 
on different components of the full wave function.

To find the quantum spectrum of (\ref{H-qu-g}) and (\ref{q-Cas-g}), we
make use of
the realization \p{bo-re-Z} for the
bosonic operators $Z^k$ and $\bar{Z}_k$, as well as
the following realization of the odd operators $\Psi^i$, $\bar\Psi_i$
\begin{equation}\label{q-re-Psi}
\Psi^i=\psi^i, \qquad \bar\Psi_i= -{\textstyle\frac{1}{2}}\,
\partial/\partial\psi^i\,,
\end{equation}
where $\psi^i$ are complex Grassmann variables. Then, the state vector
(wave function)
is defined as
\begin{equation}\label{w-f}
\Phi=A_{1}+ \psi^i B_i +\psi^i\psi_i A_{2}\,.
\end{equation}

The full wave function is subjected to the same constraints~(\ref{D0-con})
as in the
bosonic limit
(we use the normal ordering for even SU(2)--spinor operators, with all
operators
$Z^i$ standing on the right)
\begin{equation}\label{q-con}
D^0 \Phi=\bar Z_i Z^i \Phi=v^+_i\frac{\partial}{\partial
v^+_i}\,\Phi=c\,\Phi.
\end{equation}
Like in the bosonic limit, requiring the wave function $\Phi(v^+)$ to be
single-valued gives rise to the condition that
the constant $c$ must be integer, $c\in \mathbb{Z}$. We take $c$ to be
positive in
order to have a correspondence with
the bosonic limit where $c$ becomes ${\rm SU}(2)$ spin. Then \p{q-con}
implies that
the wave function $\Phi(v^+)$
is a homogeneous polynomial in $v^+_i$ of the degree $c$:
\begin{equation}\label{w-f-d}
\Phi=A^{(c)}_{1}+ \psi^i B^{(c)}_i +\psi^i\psi_i A^{(c)}_{2} \,,
\end{equation}
\begin{equation}\label{A-irred}
A^{(c)}_{i^\prime} = A_{i^\prime,}{}_{k_1\ldots k_{c}}v^{+k_1}\ldots
v^{+k_{c}} \,,
\end{equation}
\begin{equation}\label{B-irred}
B^{(c)}_i = B^{\prime(c)}_i +B^{\prime\prime(c)}_i=v^+_i B^\prime_{k_1\ldots
k_{c-1}}v^{+k_1}\ldots v^{+k_{c-1}} + B^{\prime\prime}_{(ik_1\ldots
k_{c})}v^{+k_1}\ldots
v^{+k_{c}}\,.
\end{equation}
In~(\ref{B-irred}) we extracted ${\rm SU}(2)$ irreducible parts
$B^\prime_{(k_1\ldots
k_{c-1})}$ and $ B^{\prime\prime}_{(ik_1\ldots k_{c})}$ of the component wave
functions, with the ${\rm SU}(2)$
spins $(c-1)/2$ and $(c+1)/2$, respectively.

On the physical states~(\ref{q-con}), (\ref{w-f-d})  Casimir
operator~(\ref{qu-Cas-12}) takes the value
\begin{equation}\label{qu-Cas-ev}
\mathbf{C}_2=- (c+1)^2/16 \, .
\end{equation}

On the same states, the Casimir operators  (\ref{q-Cas-3})-(\ref{q-Cas-1})
of the bosonic subgroups ${\rm SU}(1,1)$, ${\rm SU}(2)$ and ${\rm
SU}^\prime(2)$
take the following values
$$
\mathbf{T}^2=r_0(r_0-1)\,, \qquad \mathbf{J}^2=j(j+1)\,,
\qquad \mathbf{I}^2=i(i+1)
$$
For different component wave functions, the quantum numbers $r_0, j$ and
$i$ take
the values listed in the Table below.
\begin{center}
\renewcommand{\arraystretch}{1.8}
\begin{tabular}{|c|c|c|c|}
\hline & $r_0$ & $j$ & $i$ \\ \hline
     $A^{(c)}_{k^\prime}(x,v^+)$ & $\frac{c+3}{4}$ & $\frac{c}{2}$ &
$\frac{1}{2}$ \\
\hline
     $B^{\prime(c)}_{k}(x,v^+)$ & $\frac{c+3}{4} + \frac{1}{2}$ & $\frac{c}{2}
-\frac{1}{2}$ & 0 \\ \hline
     $B^{\prime\prime(c)}_{k}(x,v^+)$ & $\frac{c+3}{4} - \frac{1}{2}$ &
     $\frac{c}{2}+ \frac{1}{2}$& 0 \\ \hline
\end{tabular}\\
\end{center}
The fields $B^{\prime}_i$ and $B^{\prime\prime}_{i}$ form doublets
of SU(2) generated by $\mathbf{J}^{ik}\,$, whereas the
component fields $A_{i^\prime}=(A_{1},A_{2})$ form a doublet of ${\rm
SU}^\prime(2)$
generated by
$\mathbf{I}^{i^\prime k^\prime}$. If the super--wave function (\ref{w-f})
is bosonic
(fermionic),
the fields $A_{i^\prime}$ describe bosons (fermions), whereas the fields
$B^{\prime}_i$, $B^{\prime\prime}_{i}$
present fermions (bosons). It is easy to check that the constraint
\p{id-Cas} is
satisfied
in all cases.

Each of the component wave functions  $A_{i^\prime}$, $B^{\prime}_i$,
$B^{\prime\prime}_{i}$  carries an
infinite--dimensional unitary representation of the discrete series of the
universal
covering group of the SU(1,1)
one--dimensional conformal group. Such representations are characterized
by positive
numbers $r_0$~\cite{Barg,Per}
(for the unitary representations of SU(1,1) the constant $r_0 >0$ must be
(half)integer).
Basis functions of these representations are eigenvectors of the compact SU(1,1) generator
$$
\mathbf{R}={\textstyle\frac{1}{2}}\,\left(a^{-1}\mathbf{K}+
a\mathbf{H}\right),
$$
where $a$ is a constant of the length dimension. These eigenvalues are
$r=r_0 +n$,
$n\in \mathbb{N}$~\cite{Barg,Per,AFF}.

Using the expressions (\ref{H-qu}), (\ref{H-qu-1}), (\ref{qu-Cas-ev}) we
can write he
Hamiltonian in the form, common for all component wave functions,
\begin{equation}\label{H-com}
\mathbf{H} =\frac{1}{4}\,\left(P^2  +\frac{l(l+1)}{X^2}\right)
\end{equation}
where constant $l$ takes the values given in the Table below.
\begin{center}
\renewcommand{\arraystretch}{1.8}
\begin{tabular}{|c|c|c|c|}
\hline & $l$ \\
\hline
     $A^{(c)}_{k^\prime}(x,v^+)$ & $\frac{c}{2}$  \\ \hline
     $B^{\prime(c)}_{k}(x,v^+)$ & $\frac{c}{2} + 1$ \\ \hline
     $B^{\prime\prime(c)}_{k}(x,v^+)$ & $\frac{c}{2} - 1$ \\ \hline
\end{tabular}\\
\end{center}

Let us focus on some peculiar properties of the OSp(4$|$2) quantum mechanics
constructed.

As opposed to the ${\rm SU}(1,1|2)$ superconformal
mechanics~\cite{IKL,AIPT,Wyl},
the construction
presented here essentially uses the variables $z_i$ (or $v^+_i$)
parametrizing the
two-sphere $S^2$,
in addition to the standard (dilatonic) coordinate $x$.

Presence of additional ``spin'' $S^2$ variables in our construction leads to a
richer quantum spectrum: the relevant wave functions involve representations of the
two independent
SU(2) groups, in contrast to the ${\rm SU}(1,1|2)$ models where only one SU(2)
realized on
fermionic variables matters.

Also in a contradistinction to the previously considered models, there naturally
appears a
quantization of the conformal coupling constant which is expressed as a SU(2)
Casimir operator,
with both integer and half-integer eigenvalues. This happens already in the bosonic
sector of the model,
and is ensured by the $S^2$ variables.

\setcounter{equation}0
\section{Summary and outlook}

We have investigated ${\cal N}{=}4$ superconformal mechanics with
OSp(4$|$2) symmetry.
This model is the one--particle case (or the center-of-mass sector) of a
${\cal N}{=}4$ superconformal Calogero model recently proposed in~\cite{FIL}.
After eliminating the auxiliary and gauge degrees of freedom, we obtained
the OSp(4$|$2) generators both on the classical and on the quantum level.

The physical sector of the model is described by one ``radial'' coordinate~$x$,
four Grassmann-odd fermionic coordinates $\psi_i$ and $\bar\psi_i$ as well as
a Grassmann-even SU(2) doublet~$z_i$ which parameterizes~$S^2$.
The latter lack a standard kinetic term and appear only in a Wess-Zumino term,
i.e.~to first order in time derivatives.
These SU(2) spinor variables lead to an unusual but rather nice property:
the odd OSp(4$|$2) generators are {\it linear\/} in $\psi_i$ or $\bar\psi_i$,
as opposed to ${\rm SU}(1,1|2)$ superconformal
mechanics~\cite{IKL,AIPT,Wyl}~\footnote{
The general supergroup $D(2,1;\alpha)$ was apparently implicit in~\cite{Wyl}.}
where such generators require also terms cubic in the fermions.
Note that ${\cal N}{>}4$ supersymmetric mechanics with linear supercharges
is trivial as was indicated in~\cite{CRi}.

We observed an interesting feature which might be called a ``double
harmonic extension''. At the classical level, the worldline parameter~$t$
is extended by harmonic variables $u^\pm_i$. The above-mentioned SU(2) spinor
variables~$z^i$ can be interpreted as a kind of harmonic target variables,
in line with~\cite{ABS}. The corresponding quantum operators~$Z^i$ serve as
coordinates of a fuzzy sphere.

We performed an analysis of the quantum spectrum.
Its form relates to a subspace in the enveloping algebra of $osp(4|2)$
which is closed under the $osp(4|2)$ action.
The composite generators from this set turn out to vanish for the specific
realization of the $osp(4|2)$ superalgebra pertinent to our model.

Finally, let us discuss the links of our model to the black-hole and AdS/CFT
story. The Hamiltonian (\ref{H-com}) resembles the Hamiltonian for the radial
motion of a massive charged particle near the horizon of an extremal RN black
hole~\cite{CDKKTP} in the supersymmetry-preserving (or BPS) limit, when
the mass and electric charge of the superparticle are equal.
In our model, the quantized ``angular momentum'',\footnote{
Here it should rather be named ``SU(2) spin'' since
it can take both integer and half-integer values.}
whose square is the strength of the conformal potential, is given by
the bosonic SU(2)--spinors and is present already in the bosonic sector.
It receives corrections from the SU(2)--spinor fermions for other components
of the wave superfunction.

Despite this formal resemblance, the superconformal symmetries
differ: it is SU(1,1$|$2) for the near-horizon limit of the RN
black-hole solution of ${\cal N}{=}2, d{=}4$ supergravity, while our
model is OSp(4$|$2) invariant. Thus, one may ask to which sort of
superparticle our superconformal mechanics does correspond. This can
be explored by changing variables to the so-called AdS basis
\cite{BIK,IKN,BGIK,Gal}, in which the $d{=}1$ conformal group~${\rm
SO}(1,2)$ is realized by relativistic particle motions on
AdS${}_2\simeq{\rm SO}(1,2)/{\rm SO}(1,1)$. In addition, the
Wess-Zumino term in the action \p{4N-ph1} describes the coupling of
a charged particle on $S^2$ to a Dirac monopole in its center. 
The strength of the Wess-Zumino term is given by the product of the
electric and magnetic charges of the particle and monopole, respectively.
The potential for this magnetic flux is naturally present in the
general form of the RN solution (along with an electric potential).
However, in our case the $S^2$ variables are not propagating in either 
the conformal or the AdS bases.
Therefore, the hypothetical superparticle associated with our
superconformal model moves only on the AdS${}_2$ space and not on
the AdS${}_2\times S^2$ appearing for SU(1,1$|$2) mechanics. On the
other hand, the presence of the Wess-Zumino term suggests that our
superparticle still couples to the magnetic charge. 
It would be interesting to inquire whether a
background with such properties can arise in a black-hole type
supergravity solution in higher dimensions. 
Since the off-shell content of our model contains four bosonic degrees of 
freedom plus the worldline time for a fifth variable, we conjecture that
the appropriate supergravity should live in five spacetime dimensions.

\vfill\eject

\noindent
{\bf\large Acknowledgements}

\noindent
We acknowledge support from a grant of the Heisenberg-Landau Programme,
RFBR grants 08-02-90490, 09-02-01209, 09-01-93107 and INTAS grant 05-1000008-7928
(S.F. \&\ E.I.) and a DFG grant, project No.~436 RUS/113/669 (E.I. \&\ O.L.).
S.F. and E.I. thank the Institute of Theoretical Physics of the Leibniz University
of Hannover for the kind hospitality at the final stage of this study. S.F. thanks
M.~Vasiliev for a useful discussion.

\bigskip
\noindent
After finishing this paper, related work~\cite{BelKri} overlapping with ours
appeared in the {\tt arXiv}.

\end{document}